# First results on a laser-heated emissive probe

R. Madani,[1] C. Ionita,[2] R. Schrittwieser,[2] T. Klinger[1]

[1]*Max Planck Institute for Plasma Physics, Greifswald Branch, EURATOM Association, Germany*
[2]*Institute for Ion Physics, University of Innsbruck, Austria, e-mail: roman.schrittwieser@uibk.ac.at*
[3]*Faculty of Physics, "Al. I. Cuza" University, Iasi, Romania*

## Abstract

The floating potential $V^*_{fl,em}$ of a probe, emitting a sufficiently high electron current, yields a fairly accurate approximation of $\Phi_{pl}$. This is an advantage in comparison to the conventional Langmuir probe where, after determination of the electron temperature $T_e$, the plasma potential can only be derived indirectly from the formula $\Phi_{pl} = V_{fl} + \alpha T_e$, where $\alpha$ is a function of the ratio of the electron to the ion saturation currents ($\alpha$ is around 2.4 in a magnetised hydrogen plasma). In addition, an emissive probe also works if there are electron drifts or beams in the plasma. Emissive probes are usually realised by small directly heated loops of W-wire. Drawbacks of this design are the limited lifetime, the low electron emissivity of W and the voltage drop across the wire. We have developed a new type of emissive probe, which is heated by an infrared high-power diode laser with a maximum output power of 50 W. The probe consists of a small cylinder of $LaB_6$. The probe was inserted into the edge region of the VINETA helicon discharge plasma. Basic features of emissive probes were verified.

# 1. Introduction

For many decades, emissive probes are in use in laboratory plasmas for a fast direct determination of the plasma potential $\Phi_{pl}$. Only recently emissive probes have been applied for the first time also in the edge plasma region of fusion experiments for measuring $\Phi_{pl}$ and related parameters like the electric field and fluctuations of the potential and electric field [1,2,3,4,5]. The floating potential $V_{fl,em}^*$ of a probe, which emits a sufficiently high electron current, is close to $\Phi_{pl}$. In contrast to that, with conventional Langmuir probes the plasma potential can only be determined indirectly from the formula

$$\Phi_{pl} = V_{fl} + \alpha T_e = V_{fl} + T_e \ln\left(\frac{I_{es}}{I_{is}}\right), \quad (1)$$

where $I_{es,is}$ are the electron and ion saturation currents, respectively. This formula can be derived from simple Langmuir probe theory for the probe voltage $V_p \leq \Phi_{pl}$. Obviously, for this formula $T_e$ has to be determined at first, which is not an easy task, especially in the edge region of a toroidal plasma, where there can be strong gradients and fluctuations of the temperature. The factor $\alpha = \ln(I_{es}/I_{is})$ is in general around 2.4 for hydrogen in a magnetised plasma [2]. For an emissive probe Eq. (1) becomes

$$\Phi_{pl} = V_{fl,em} + T_e \ln\left(\frac{I_{es}}{I_{is} + I_{em}}\right), \quad (2)$$

where $I_{em}$ is the current emitted from the probe into the plasma [2]. For $I_{em} = I_{es} - I_{is}$, the electron current from the plasma is compensated and the floating potential of the probe is shifted completely towards $\Phi_{pl}$, i.e., $V_{fl,em} \equiv V_{fl,em}^* = \Phi_{pl}$. Eqs. (1,2) are, as mentioned above, only valid for $V_p \leq \Phi_{pl}$, purely Maxwellian electrons and for neglecting space charges around the probe. For comparison with the experimental data it is practical to define the difference between the actual floating potential and the plasma potential normalised to the electron temperature:

$$\Delta \equiv \frac{\Phi_{pl} - V_{fl,em}}{T_e} \quad (3)$$

For our assumption of a purely Maxwellian plasma, $\Delta$ is equal to:

$$\Delta = \ln\left(\frac{I_{es}}{I_{is} + I_{em}}\right) \quad (4)$$

Eqs. (3,4) demonstrate the typical behaviour of an emissive probe: the more it is heated, the larger is $I_{em}$, the smaller $\Delta$ becomes and the more its floating potential approaches $\Phi_{pl}$, un-til for $I_{em} = I_{es} - I_{is}$, $\Delta = 0$.

For a cold probe these equartions reduce to

$$\Delta_0 \equiv \frac{\Phi_{pl} - V_{fl}}{T_e} = \ln\left(\frac{I_{es}}{I_{is}}\right) = \alpha \qquad (5)$$

A typical conventional emissive probe is realised by a loop of tungsten wire of about 0.2 mm diameter and about 8 mm length, which is inserted into a double-bore ceramic tube. The probe is heated by an external power supply or battery until electron emission starts. Drawbacks of this design are the limited lifetime of the tungsten wire, the low electron emissivity of tungsten and the voltage drop across the wire.

## 2. Probe design

We have developed a prototype of a new emissive probe, which is heated by an infrared high-power diode laser JenLas HDL50F from the company JenOptik, Jena, Germany, with a maximum output power of 50 W and a wavelength of 808 nm. The laser beam is coupled to a conventional glass fibre of about 3 m length that terminates in a lens head, by which, in a distance of 15 cm, a focus of 0.6 mm diameter can be produced. This laser-head was positioned directly on a quartz-glass window perpendicular to the direction of the probe insertion. Fig. 1 shows this set-up schematically.

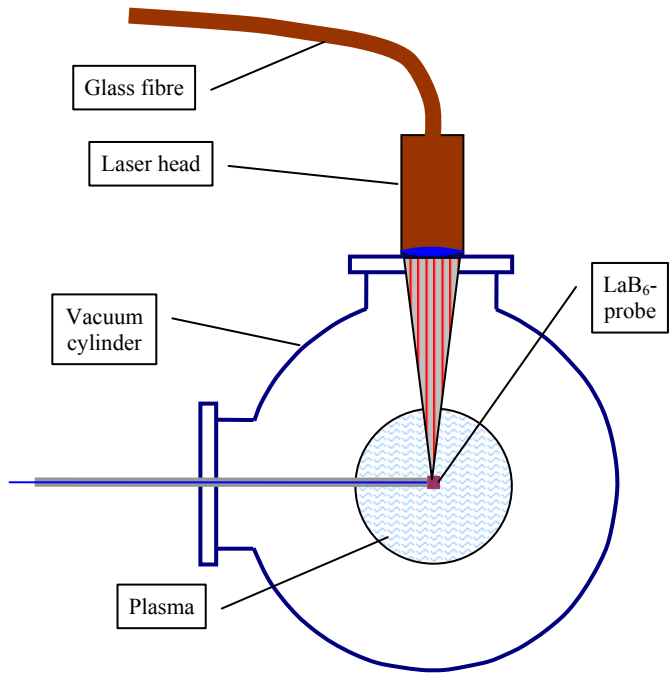

*Fig. 1. Schematic of the laser-heated probe*

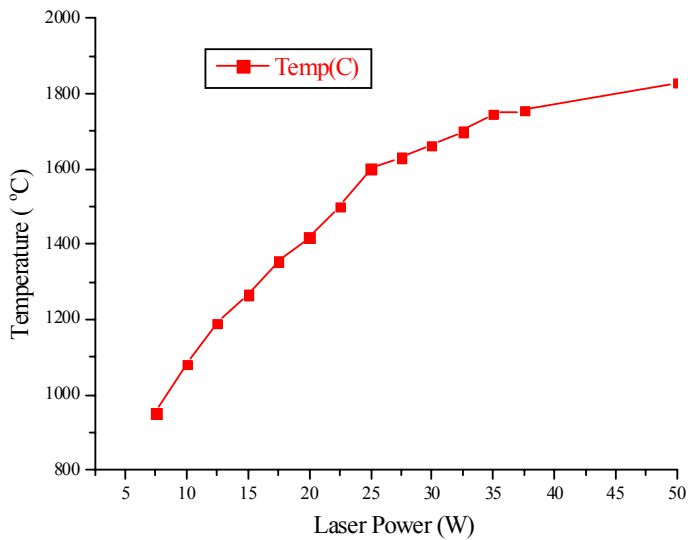

*Fig. 2. Increase of the temperature of the LaB$_6$ probe with the laser power.*

The probe consists of a small cylinder of lanthanum hexaboride (LaB$_6$) with a diameter of 3.2 mm and a height of 2.2 mm. This gives a total area of about $A_p \cong 38$ mm$^2$. The LaB$_6$ electrode is connected to a molybdenum wire of 0.2 mm diameter, which also provides the electrical connection to the probe. The Mo-wire was spliced with a number of copper threads and pulled through a one-bore ceramic tube [6]. The probe is inserted into the plasma cylinder of VINETA [7], which produces a magnetized argon plasma of 10 cm diameter and 4 m length in a magnetic field up to 0.1 T with a density of about $10^{19}$ m$^{-3}$, an electron temperature of 3 eV and an ion temperature of 0.2 eV. The plasma is produced by a helicon discharge at moderate radiofrequency powers of less than 6 kW.

## 3. Experimental results and discussion

Fig. 2 shows the increase of the temperature of the LaB$_6$ probe with the laser power, which was determined without plasma. For 50 W the temperature of the LaB$_6$ probe piece reaches 1830°C, at which value, according to the Richardson law (with the Richardson constant $A^*_{LaB_6} = 29$ A/cm$^2$K$^2$ and the work function $W_{LaB_6} = 2,66$ eV), this substance emits an electron current density of about 54 A/cm$^2$ under field-free emission conditions. According to that, from our probe with $A_p \cong 38$ mm$^2$ we can in principle produce an emission current of about 20 A.

Fig. 3 shows a set of current-voltage (*I-V*) characteristics of the laser-heated emissive probe in the plasma with increasing laser heating power. The typical behaviour of a probe with increasing electron emission current is well verified. At first, when the probe is not heated at all (black line), the *I-V* trace is that of a cold probe with a very small ion saturation current on the left-hand side and a much larger electron saturation current on the right-hand side. We notice that the floating potential of the probe is on the negative side, i.e., $V_{fl} \cong -3.3$ V. When the laser heating is turned on and the temperature of the probe increases, an electron emission current is produced which superimposes on the ion current. The higher the emission current becomes, the more the floating potential shifts to the right-hand side until above about 20 W a saturation of this value occurs at $V^*_{fl,em} \cong +8.8$ V. This is in agreement with the well-known behaviour of an emissive probe [2], and usually $V^*_{fl,em}$ is taken as a sufficiently good approximation for $\Phi_{pl}$.

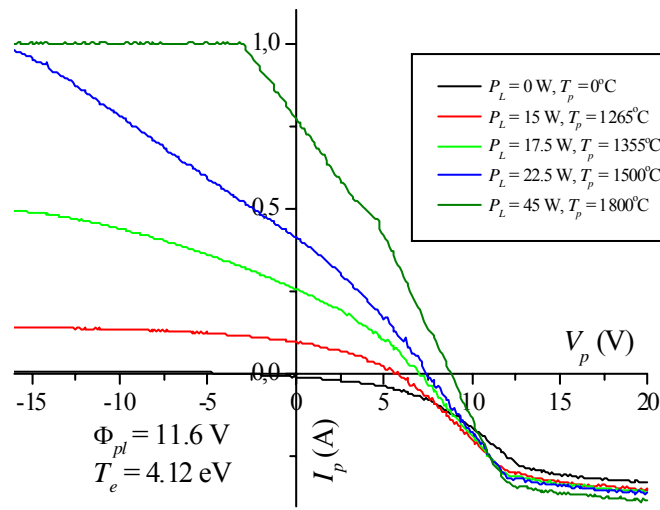

*Fig. 3. I-V characteristic of the probe with the laser power and the temperature of the LaB$_6$ probe as parameter, respectively. The indicated electron temperature and plasma potential were determined from the cold characteristic (black curve) for $P_L = 0$ W.*

However, the inserted value of the plasma potential, $\Phi_{pl} = +11.6$ V, (and of the electron temperature, $T_e = 4.12$ eV) have been determined from the cold $I$-$V$ characteristic (for $P_L = 0$ W) for comparison. Immediately we see what is also visible for the eye from the characteristics of Fig. 3, namely that the saturated value of the emissive probe floating potential $V^*_{fl,em} \cong +8.8$ V remains considerably below this value.

One more surprising fact, which is in contradiction to the usual idealised concept of an emissive probe [8], can be seen from the $I$-$V$ traces of Fig. 3: With increasing emission current, also the electron saturation current from the plasma to the probe increases somewhat. Also this fact has already been observed previously [2], but it is not yet clarified satisfactorily.

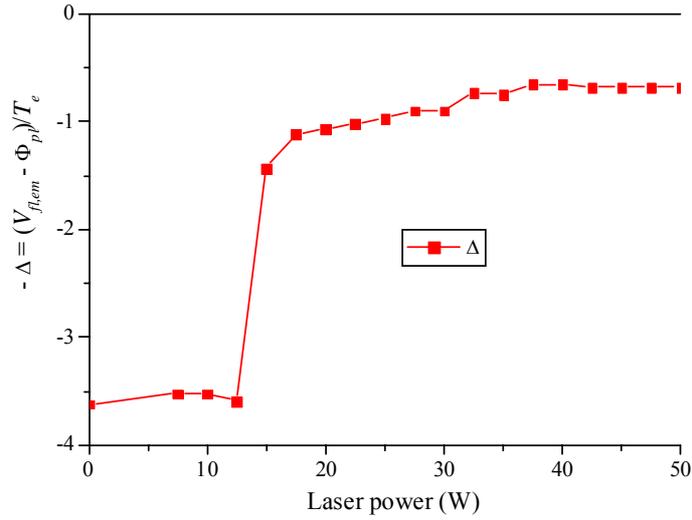

Fig. 4. Negative difference between the actual floating potential of the probe and the plasma potential determined from the cold characteristic, normalised to the electron temperature (Eq. 3) versus laser heating power. Here the negative value of Δ is shown for greater lucidity.

Fig. 4 shows the variation of the floating potential of the probe with the laser heating power, here demonstrated by the negative difference $-\Delta$ (Eq. 4) between the actual value $V_{fl,em}(P_L)$ and $\Phi_{pl} = 11.6$ V. For better insight the negative value of Δ is shown here since it indicates the shift of $V_{fl,em}(P_L)$ towards the plasma potential. Starting from a value of $\Delta \cong 3.6$ (which corresponds to the cold floating potential $V_{fl} \cong -3.3$ V) of the unheated probe, with increasing heating power the floating potential rises. At a laser power of $P_L \cong 13$ W, $V_{fl,em}$ jumps up and reaches the aforementioned final saturation of $V^*_{fl,em} = +8.8$ V for $P_L \geq 40$ W, which corresponds to $\Delta \cong 0.68$. This value lies about 2.9 $T_e$ above $V_{fl}$, thus $\Delta_0 \cong 2.9$ (Eq. 5).

As already indicated above, we observe that the saturated value of $V^*_{fl,em}$ remains by about 0.68 $T_e$ below the value of $\Phi_{pl} = +11.6$ V determined from the cold characteristic. This systematic discrepancy is well known [9,10], and the value 0.68 $T_e$ is in good agreement with an emissive probe simulation by Reinmüller [11]. It is ascribed to the formation of a space charge sheath around the probe consisting of emitted electrons, which cannot leave the probe even for $V_p \leq \Phi_{pl}$ due their much lower temperature $T_p \cong 0.2$ eV than that of the plasma electrons ($T_e = 4.12$ eV). Therefore the average unperturbed current density of plasma electrons is higher than that of emitted electrons. The probe, having to re-establish the necessary equilibrium for the floating condition between the two opposite electrons fluxes towards and away from it, makes its floating potential lower than $\Phi_{pl}$. Thereby the flux of plasma electrons towards the probe is reduced whereas the emitted electrons are accelerated towards the plasma until their fluxes are equal again. Compared to these fluxes, the ion flux is so small that we have ignored it for our basic considerations.

We note, however, that Ye and Takamura [9] found that the floating potential of a strongly emitting probe should be around one $T_e$ below the real value of the plasma potential. This is not in keeping with our above-mentioned result.

Also the slight increase of the electron saturation current on the right side of the characteristic with increasing emission current is perhaps due the emitted electrons. Although on this side, where $V_p > \Phi_{pl}$, the emitted electrons cannot leave the probe, they might still expand the electron-rich space charge sheath around the probe, thereby increasing the probe area which is effective for electron collection. In this case the space charge is of course electron-rich and primarily formed by the plasma electrons which are accelerated towards the probe. This effect might have something in common with the formation of an ion-rich space charge sheath in front of a positively biased plane probe in a DP-machine plasma, in which a strong ion beam is produced from the source chamber towards the probe [12], by which effect the effective collecting area of the probe is also strongly increased and an additional knee is found in the I-V characteristic of the probe.

Both above-mentioned effects that are possibly related to a space charge of emitted electrons need further clarification. To this end, the diode system consisting of the emissive surface of the probe and the surrounding plasma has to be treated comprehensively. At the moment, simulations approaches are under way, which treat this system in various ways.

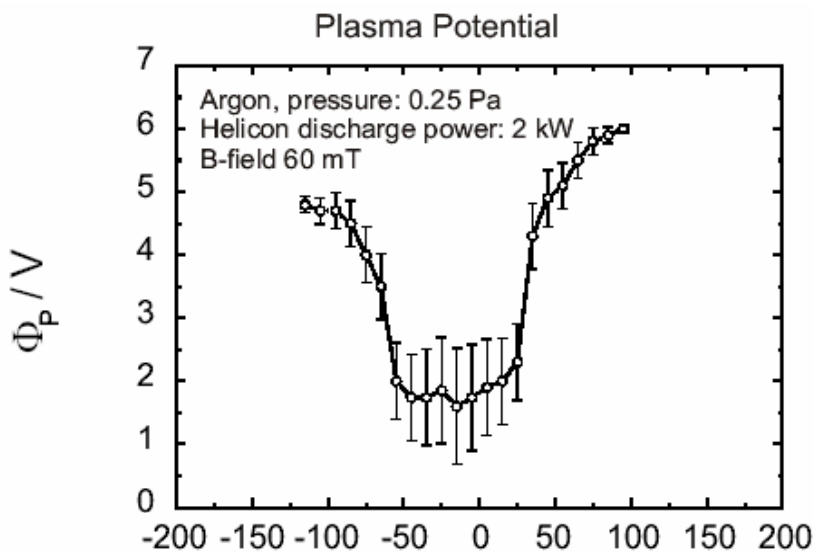

*Fig. 5. Radial profile of the plasma potential through the VINETA plasma column, measured with the laser-heated emissive probe.*

Fig. 5 shows a radial profile of the plasma potential through the VINETA plasma column, measured by the laser-heated emissive probe. These results are in keeping with analogous measurements in a Q-machine, but do not agree with other measurements in VINETA. Also this discrepancy has still to be clarified.

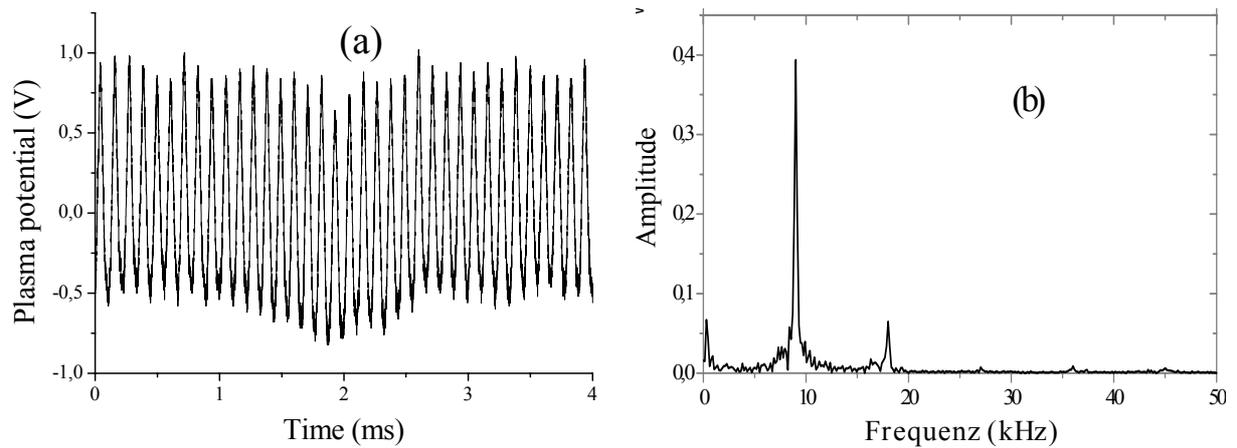

*Fig. 6. (a) Times series of a plasma potential signal taken by the emissive probe in the edge region of the VINETA plasma cylinder. (b) Amplitude FFT of the same signal.*

Fig. 6 shows a typical regular oscillation signal which has been registered with the probe fully heated emissive probe. The position was in the edge region of the VINETA plasma column in the helicon regime at an argon pressure of 0.22 Pa. Fig. 6a shows the time series of the oscillations, from which we discern that the oscillation occurs between about –0.5 V and +1.0 V, thus has an amplitude of 0.75 V approximately. Fig. 6b shows the FFT of the same signal. We see that the first harmonic has a frequency of about 9 kHz. The most probable explanation for this instability is a drift type one, driven by the density gradient in the edge region.

## 4. Conclusion

We have succeeded to construct a laser-heated electron emissive probe which can produce a much higher emission current than a conventional emissive wire probe. It has also a much longer life time since we have observed no evaporation or sputtering of the $LaB_6$ piece even after many hours of constant strong irradiation with the infrared laser. The probe has also a better time response since no electric heating system with a high internal capacity is necessary.

**Acknowledgements:** This work has been carried out within the Association EURATOM-IPP and EURATOM-ÖAW. The content of the publication is the sole responsibility of its author(s) and it does not necessarily represent the views of the Commission or its services. The support by the Fonds zur Förderung der wissenschaftlichen Forschung (Austria) under grant No. P-14545 is acknowledged.